\documentclass[epj,twocolumn]{webofc}
\usepackage[varg]{txfonts}   
\usepackage{graphicx}

\wocname{epj}
\woctitle{Seismology of the Sun and the Distant Stars 2016}
\begin{document}
\title{Presence of mixed modes in red giants in binary systems}
%

\author{\firstname{Nathalie} \lastname{Theme\ss l}\inst{1,2,3}\fnsep\thanks{\email{themessl@mps.mpg.de}} \and
        \firstname{Saskia} \lastname{Hekker}\inst{1,2}
        \and
        \firstname{Yvonne} \lastname{Elsworth}\inst{4,2}
}

\institute{Max Planck Institute for Solar System Research, Justus-von-Liebig-Weg 3, 37077 G{\"o}ttingen, DE
\and
           Stellar Astrophysics Centre, Department of Physics and Astronomy, Aarhus University, Ny Munkegade 120, 8000 Aarhus C, DK
\and
           Institute for Astrophysics, Georg-August University G{\"o}ttingen, Friedrich-Hund-Platz 1, 37077 G{\"o}ttingen, DE
\and
		School of Physics and Astronomy, U. Birmingham, Edgbaston, Birmingham B15 2TT, UK
          }

\abstract{%
The frequencies of oscillation modes in stars contain valuable information about the stellar properties. In red giants the frequency spectrum also contains mixed modes, with both pressure (p) and gravity (g) as restoring force, which are key to understanding the physical conditions in the stellar core. We observe a high fraction of red giants in binary systems, for which g-dominated mixed modes are not pronounced. This trend leads us to investigate whether this is specific for binary systems or a more general feature. We do so by comparing the fraction of stars with only p-dominated mixed modes in binaries and in a larger set of stars from the APOKASC sample. We find only p-dominated mixed modes in about 50\% of red giants in detached eclipsing binaries compared to about 4\% in the large sample. This could indicate that this phenomenon is tightly related to binarity and that the binary fraction in the APOKASC sample is about 8\%.
}
\maketitle
%
\section{Introduction}
\label{intro}
In its nominal mission the {\it Kepler} space telescope \cite{2010bor} provided long-term photometric data for more than 100\,000 stars, among which many pulsating red giants were found. Red giants are evolved stars showing solar-like oscillations which are stochastically excited and intrinsically damped by the near surface convection. Their oscillation spectrum comprises several overtones of radial order ($n$) and spherical degree ($\ell$) modes. The most significant peaks form a well-defined pattern which reveals the structure of the radial ($\ell=0$) and non-radial ($\ell=1,2$) modes \citep{1980tas,2011mos}. Radial modes are pure pressure modes that are mostly sensitive to the outer layers of the star. In contrast, non-radial modes have a mixed nature with gravity as the restoring force in the deep interior and pressure as the restoring force close to the stellar surface. The dipole ($\ell=1$) modes consist of one central most p-dominated mode, which is surrounded by several mixed modes that have a more g-dominated character. In a significant number of red giants in detached binary systems these g-dominated mixed modes are missing. In close binary systems, tidal interactions can impact upon the oscillation spectra of the individual stars in the form of mode damping or even complete mode suppression \cite{2013gau}. This interplay between binary components is not expected in detached systems, where stars are assumed to evolve separately. In our study, we investigate if the lack of pronounced g-dominated mixed modes is mainly observed in red giants that belong to binary or triple-star systems or if it is also a common phenomenon among field red giants.

\section{Data}
\label{sec-1}

\subsection{Binary and triple-star systems with red-giant components}
\label{subsec-1}
We used {\it Kepler} long-cadence ($\sim29.4$\,min) data for a sample of red giants that were confirmed to be components of binary or triple-star systems \cite{2013gau,2014bec,2014gau,2016gau}. For this study, we adopted the unweighted power density spectra of the corrected and concatenated timeseries which are provided by KASOC \cite{2014han}.

\subsection{APOKASC red giants}
\label{subsec-2}
For comparison, we chose 6604 red giants from the APOKASC \cite{2014pin} sample, which contains many astrophysically interesting targets whose stellar properties are available from asteroseismology and complementary spectroscopy. We adopted the corrected lightcurves \cite{2014han} that are available in the APOKASC data bundle to calculate the power density spectra of the red-giant stars. These data are the same as used by \cite{2017els}.

\section{Method}
\label{sec-2}
\begin{figure*}[h]
\centering
\includegraphics[height=\hsize,clip,angle=90]{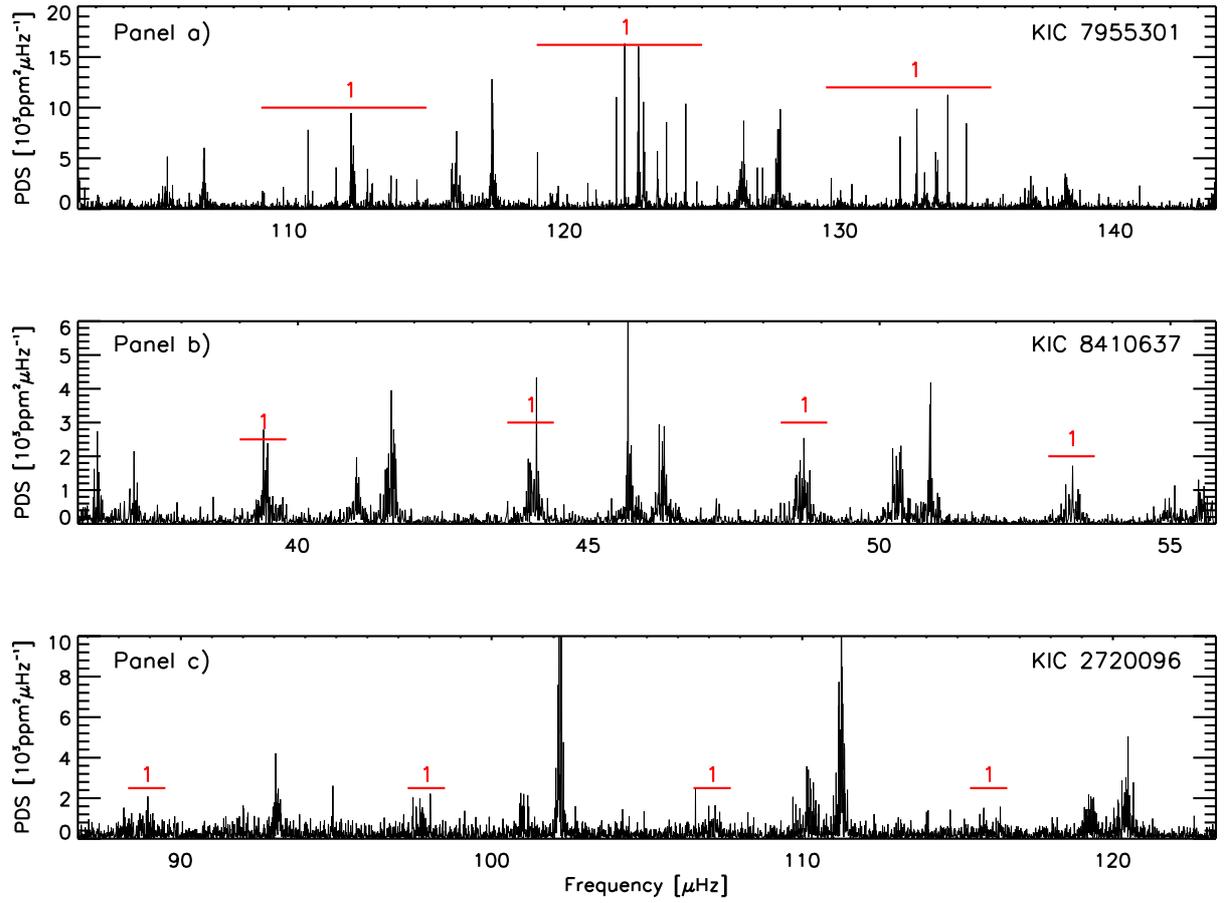}
\caption{The frequency range of oscillations around the frequency of maximum oscillation power. Panel a) shows KIC\,7955301 which exhibits clear mixed modes. Panel b) shows KIC\,8410637 where only the p-dominated non-radial modes are distinct. Panel c) shows KIC\,2720096, a red giant with suppressed dipole modes.}
\label{fig-1}       
\end{figure*}

In this study, we attempt to classify the red giants into three different categories: (a) stars where g-dominated mixed modes are clearly present, (b) stars that show only p-dominated dipole modes, and (c) stars exhibiting non-radial mode suppression. See also Figure\,\ref{fig-1} for an example of each of these classes.

For the characterization of the red giants, we performed a visual inspection of the dipole $\ell=1$ modes in the oscillation region close to the frequency of maximum oscillation power ($\nu_{\rm max}$). Thus, we restricted the power density spectrum of each star to $\pm\,2.5\,\Delta\nu$ around $\nu_{\rm max}$, where $\Delta\nu$ is the large frequency separation between modes of the same spherical degree and consecutive radial orders. Based on this approach, we inspected 18 red giants in binary and triple-star systems and we extended this classification to a larger set of stars from the APOKASC sample. The latter contains red giants that cover different evolutionary stages. We were not able to categorize 284 of these stars either due to very low signal-to-noise spectra, oscillation regions that were too close to the Nyquist frequency or the absence of solar-like oscillations. Table~\ref{tab-1} presents the preliminary results with the percentage of red giants according to the three different types that we defined. 

\begin{table*}
\centering
\caption{The percentage of stars in the different categories.}
\label{tab-1}       
\begin{tabular}{lccccc}
\hline
& Number  & Mixed & P-dominated & Suppressed & Unidentified   \\
& of stars  & modes & modes & modes &  \\\hline
Binaries & 18  & $\sim39$\% & $\sim50$\% & $\sim11$\% & 0\% \\\hline
APOKASC &    6604 & $\sim85$\% & $\sim4$\% & $\sim7$\,\% & $\sim4$\%\\\hline
\end{tabular}
\end{table*}

\section{Discussions}
\label{sec-dis}
In this preliminary study, we found a high percentage ($\sim50$\%) of red giants in known detached binary systems that show mainly p-dominated dipole modes. To investigate the significance of this result, we also examined a large number of red giants from the APOKASC sample. \cite{2017els} detected distinct mixed dipole modes in a large fraction of these stars. We found that the remaining small fraction of stars show either p-dominated or suppressed dipole modes. Most interestingly, the fraction of stars exhibiting p-dominated dipole modes is much smaller than what we observed for the binaries, where about half of the stars showed a tendency towards this category. In contrast, the fraction of suppressed dipole modes is of the same order of magnitude for both samples. As we use a statistically insignificant number of known binaries, we are not able to draw any firm conclusions from this. We can however speculate that the binary nature does not seem to be the cause for mode suppression. Yet binarity appears to have some influence on the g-dominated mixed modes, which we will investigate further. If we take this one step further, saying that the presence of mainly p-dominated mixed modes only appears in about half the binaries, this would mean that the binary fraction in the APOKASC sample would be $\sim8$\%.

\section*{Acknowledgements}
NT and SH have received funding from the European Research Council under
the European Communitys Seventh Framework Programme (FP7/2007-2013) / ERC grant agreement no 338251 (StellarAges). NT also acknowledges travel support from the EC Project SPACEINN (FP7-SPACE-2012-312844). YE acknowledges the support of the UK Science and Technology Facilities Council (STFC). Funding for the Stellar Astrophysics Centre (SAC) is provided by The Danish National Research Foundation (Grant agreement no.: DNRF106).\\

 \bibliography{themessl}
%
%
%
%

\end{document}